\documentclass{article}
\usepackage{graphicx}
\usepackage{amsmath, amsthm, amssymb}
\usepackage{url}
\usepackage[backend=biber, bibencoding=utf8, style=ieee]{biblatex}
\usepackage{xcolor}   
\usepackage{booktabs} 
\usepackage[margin=2.5cm]{geometry} 
\usepackage [english]{babel}
\usepackage [autostyle, english = american]{csquotes}
\MakeOuterQuote{"}
\usepackage{authblk}

\begin{document}

\title{Heat transfer correlations for buoyant liquid metal MHD flows in blanket poloidal channels }


\author[1]{Daniel Suarez\thanks{Corresponding author: daniel.suarez.cambra@upc.edu}}
\author[2]{Elisabet Mas de les Valls}
\author[1]{Llu\'is Batet}

\affil[1]{Department of Physics, Universitat Polit\`ecnica de Catalunya (UPC)}
\affil[2]{Heat Engines Department, Universitat Polit\`ecnica de Catalunya (UPC)}

\maketitle

\begin{abstract}
In recent years, several simulation codes for reproducing liquid metal magnetohydrodynamic (MHD) phenomena have been validated and benchmarked.
Accurate simulation codes are crucial to enhance our understanding of how flow behavior affects heat transport in liquid metal-based breeding blankets. 
Using heat transfer correlations, that model the influence of flow characteristics on the transport of heat, is especially interesting for system designers because it saves them the effort and time in completely simulating every design proposal.
Our group has studied the buoyant MHD flow in poloidal channels on the EU Dual Coolant Lead Lithium (DCLL) blanket geometry.
Two different codes were used for this study: a 2D fully-developed code and a Q2D-fully-developed code.
In this work, we explored the influence of different flow conditions in the heat transport phenomena parametrically. 
This article presents the results of the calculations performed using the two codes and provides heat transfer correlations for poloidal EU DCLL channels.\end{abstract}

\section{Introduction}
\label{section:Intro}

Dual-coolant lead-lithium (DCLL) is a breeding blanket design that is being considered within EU DEMO project as advanced blanket \cite{federici_2019}. 
The latest DCLL design (provided by CIEMAT (\cite{ivan2019} and \cite{SMS}) consists of a single-module segment (SMS), avoiding U-turns in the liquid metal flow. The walls containing the PbLi flow will be of a ceramic insulating material, avoiding the electric coupling between fluid and walls.

\begin{figure}[!h]
  \centering
   \includegraphics[width=0.85\textwidth]{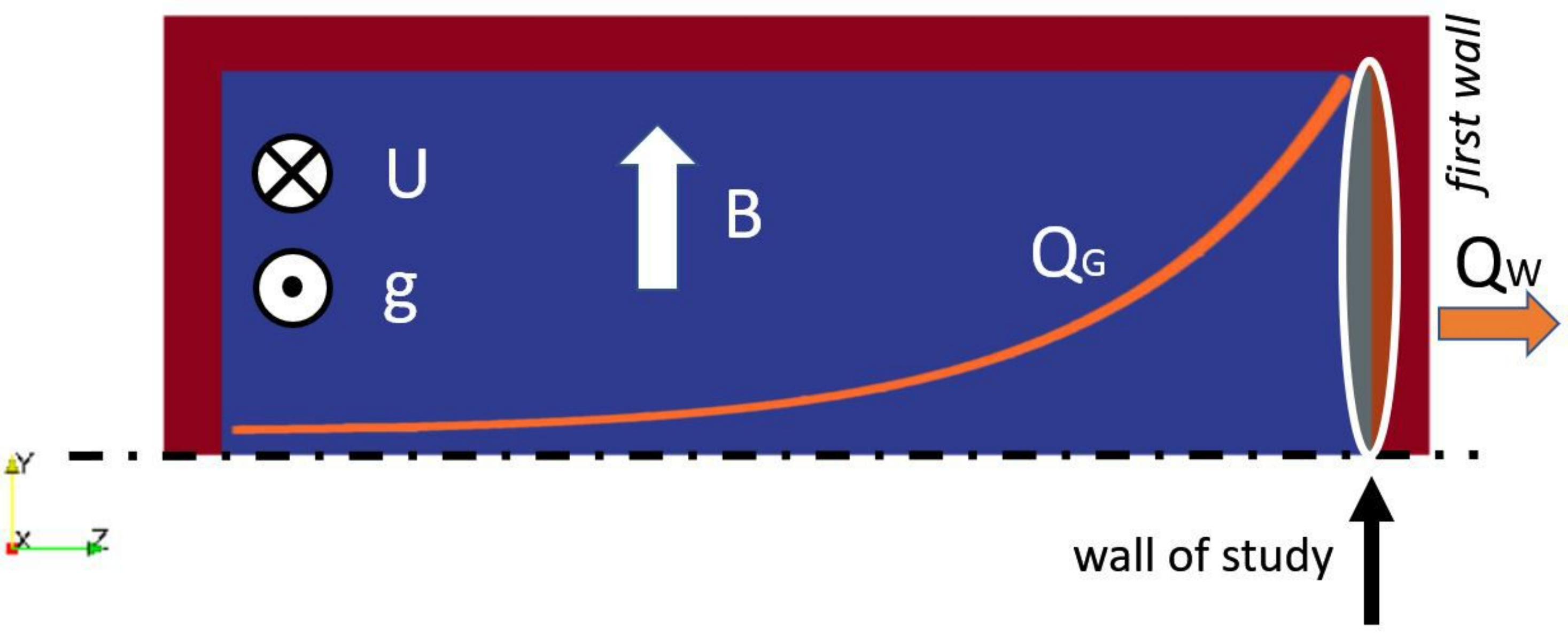}
   \caption{Wall of study \cite{Suarez2021}}
   \label{fig:wallofstudy}
\end{figure}

In a recent work \cite{Suarez2021}, our group explored the 2D fully-developed analysis of heat transfer in buoyant liquid metal MHD flows in DCLL outboard channels. 
The analysis used an electric potential MHD model to study the heat transfer coefficient applicable to a specific wall, as shown in Figure \ref{fig:wallofstudy}.

$U$ represents the velocity direction (perpendicular to the studied plane), $g$ is the gravity vector, opposed to $U$, $B$ is the magnetic field direction, $Q_G$ represents the volumetric exponential heat generation, and $Q_W$ is the heat extracted through the outer wall by the helium cooling channels of the first wall.
The rest of the walls are adiabatic. 
Note that the "wall of study" is the plane that separates the fluid and solid regions.

The work presented here is a continuation of that started in 2020 \cite{Suarez2021}. Our group had access to High Performance Computer resources and was capable to implement a much deeper parametric study on the influence of several parameters to the Nusselt number, characteristic of the heat transfer phenomena.

More specifically, the variables studied in this work are: the aspect ratio of the channel ($AR$), the wall conductance ratio ($c_W$), the Grashof number ($Gr$), the Grashof ratio ($GrR$), the Hartmann number ($H\!a$), the volumetric heat generation shape coefficient ($m$), and the Reynolds number ($Re$). The influence of all of them to the Nusselt number is analyzed.

Recently, our group has worked on the development and validation of a quasi-two-dimensional (Q2D) MHD model \cite{Suarez2023}. 
In this study, we compare the results obtained by the electric potential MHD 2D model with the 1D fully-developed Q2D model, as described below.

\section{Methodology}
\label{section:Methodology}

The definition of the flow conditions is done using the dimensionless numbers of each field: the Hartmann number for the magnetic field intensity ($H\!a=B b \sqrt{\sigma_m/\mu}$), the Reynolds number for the velocity field ($Re=\rho U D_h /\mu$),  the wall conductance ratio for the wall conductivity ($c_w=\sigma_w t_w /\sigma_m b$), the Grashof number for the heat generated in the fluid domain ($Gr=g\beta \Delta T a^3 /\nu^2$), and the Grashof ratio for the extracted heat through the outer wall ($GrR = Q_W/Q_G$). Parameters $\sigma_w$ and $t_w$ correspond to wall electric conductivity and wall thickness, $a$ is half width of the channel in the heat flux radial direction, $b$ is half width of the channel in the magnetic field direction and $D_h$ is the hydraulic diameter. The aspect ratio, also studied, is defined as $AR=a/b$.

The temperature difference $\Delta T$ is provoked by $Q_G$, a non-uniform volumetric heating (only in the liquid metal) caused by neutron flux interaction as described in \cite{Garcinuno2018}. The wall region has no volumetric heating. By convention, the $Gr$ number definition determines the $\Delta T$, and it is used to dimension the mean volumetric heat generation with $\overline{q}=k\cdot \Delta T / a^2$, with $k$ the liquid metal thermal conductivity. Total heat generation is $Q_G=\overline{q}\cdot V$. The heat deposition is modelled as an exponential profile shape, then $Q_G=\int_{V} q_0 \cdot e^{-mz} \cdot dV$. Finding $q_0$ allows to find the source term:

\begin{equation}
    \label{eq:exponential}
    S_{th}=S_0 \cdot \mathrm e ^{-m\cdot z}
\end{equation}
where $S_0=q_0/(\rho \cdot C_p)$. $z=0$ is the centre of the channel width in the $z$ direction as shown in Figure \ref{fig:wallofstudy}. $m$ is the volumetric heat generation shape coefficient, mentioned at the end of Section \ref{section:Intro}.

The first code used for this analysis is based in the electric potential as main electromagnetic variable: due to the negligible induced magnetic field, the low magnetic Reynolds number formulation is suitable. 
The buoyant effect is modelled in the momentum equation using the Boussinesq approximation. 
A detailed derivation of the applicable set of equations (described in \cite{Suarez2021}) and the PISO-like solution procedure using a finite volume approach can be found on the code development description by Mas de les Valls \cite{elimas}.
The code is capable to solve multiple regions, guaranteeing the conservation of electric current density at the walls, as showed a recent validation and verification exercise \cite{Suarez2022}.
In this work, this code will be referred as \textit{"Epot"}, for electric potential.
All the details associated with the physical properties of the materials, the mesh and the discretization strategy can be found in \cite{Suarez2021}.

\begin{figure}[!h]
  \centering
   \includegraphics[width=0.85\textwidth]{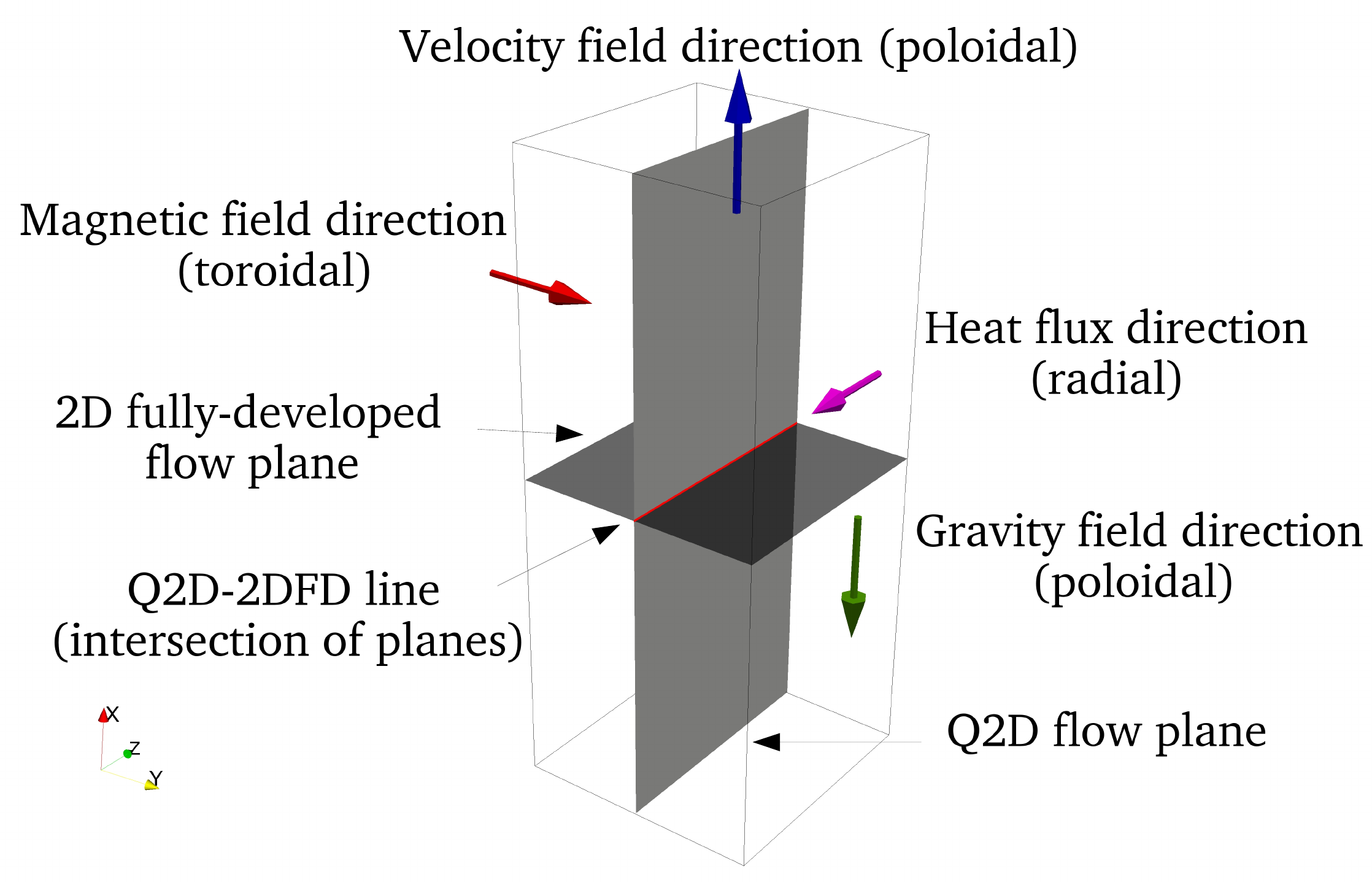}
   \caption{Schematic of a liquid metal MHD flow channel with the planes that allow simplification of the governing equations}
   \label{fig:planes}
\end{figure}

The second code used for this analysis is based in the Q2D MHD model proposed by Sommeria and Moreau in 1982 \cite{Sommeria1982}.
The model simplifies the flow - electromagnetic interaction adding a single term to the momentum equation.
A recent work published by our group \cite{Suarez2023} describes the implementation and validation of the model.
The Q2D model considers the hypothesis that for a strong enough magnetic field the influence of the Hartmann layers is sufficiently small to consider that the flow behaves as 2D, within the plane perpendicular to the magnetic field. 
The heat transfer configuration model of a fully-developed flow in the poloidal channels studied here is specially suitable for the use of the Q2D model. 
The ceramic walls are electrically insulated, what is a requirement for using Q2D approach, and the flow is fully-developed, what means that only 1D needs to be calculated, reducing drastically the computational time.
Figure \ref{fig:planes} shows the planes where the flow model can be simplified and their intersection line, for a liquid metal MHD flow channel with a transverse magnetic field and transverse heat flux.

Using the 1D model implies that the results associated with the Q2D code ignore the temperature distribution in the same plane, as for example the heat extracted through the Hartmann layer or around the channel corners.
This can lead, of course, to different results compared to the 2D model calculated with the electric potential approach. 
In this work we will assess the adequacy of using the Q2D model in this configuration.

All simulations have been carried out using a second order linear discretization scheme and the steady state criterion was set to $max((\psi-\psi_0)/\psi_0) < 10^{-6}$, with $\psi$ either velocity or temperature and $\psi_0$ the previous time step value. 

The Nusselt number, is defined as $N\!u=h\cdot a / k$ with $h$ the heat transfer coefficient calculated as $h=Q_{bw}/(A_{bw}\cdot \Delta T_{bw})$. $\Delta T_{bw}$ is defined as $\Delta T_{bw} = \overline{T}_b-\overline{T}_{bw} = \overline{\theta}_{bw}$. 
$A_{bw}$ is the area of the interface between bulk and wall in the "wall of study", that can be seen in Figure \ref{fig:wallofstudy}.
$\overline{T}_{b}$ is the mean temperature in the bulk 2D plane, $\overline{T}_{bw}$ is the mean temperature of the "wall of study" and $Q_{bw}$ is the heat flux through the same surface (refer to Figure \ref{fig:wallofstudy}): $Q_{bw} = \int_{f} k \cdot \nabla \theta_f \cdot \delta A_f$. 
The area of each discretized face is $A_f$ and $\nabla \theta_f$ is the temperature gradient in such face. 
Nusselt number is dimensionless.

\section{Base case and selected range of parameters}

The parametric study is conducted using a base case, from which each variable is modified independently. The selected base case\footnote{When preparing the cases to be studied in this work, we realized that the definition of the Hartmann number in the previous article \cite{Suarez2021} was incorrect. To keep the adequacy of the values presented in that work, the real definition must have been $H\!a=B 2 b \sqrt{\sigma_m/\mu}$, note the "$2$". Otherwise, using the most common definition of $H\!a=B b \sqrt{\sigma_m/\mu}$, the range of values starts from $H\!a=1500$. Similarly, a later energy balance within the studied region showed that when the $GrR$ was intended to be $GrR=0.02$ in \cite{Suarez2021}, it was actually $GrR=0.00938$. All these was corrected for this work.}, together with the minimum and maximum values of all variables are presented in Table \ref{tab:cases}.

\begin{table}[!h]
    \centering
    \caption{Selected set of dimensionless variables for the parametric analysis}
    \begin{tabular}{l|c c c }
    \toprule
              &  Min.       &   Base case  &  Max.    \\
    \midrule
        $H\!a$  &  1500      &   1500    &   10000    \\
        $Re$  &  1000      &   3000    &   20000    \\
        $Gr$  &  1e6       &   1e7     &   1e8      \\
        $GrR$ &  0.00023    &   0.00938    &   0.01407     \\
        $c_w$ &  1e-16     &   1e-12   &   1        \\
        $AR$  &  0.25      &   1.5     &   3        \\
        $m$   &  1         &   6.3     &   10       \\
    \bottomrule
    \end{tabular}

    \label{tab:cases}
\end{table}

The selection of the parameters has taken into account the findings presented by Vetcha et. al. \cite{Vetcha2013} in 2013, showing how of the combination of $Re$, $H\!a$ and $Gr$ numbers influence on the stability of a volumetrically heated rectangular vertical duct, with ascending flow.
It proposed that above certain critical $H\!a_{cr}$ number (in combination with a $Re$ and $Gr$ numbers), the flow is linearly stable.
Although the range of applicability ($10^6<Gr<10^9$) is far from the proposed operational conditions ($Gr \sim 10^{12}$), making a simple extrapolation, one could suggest that in such conditions the flow is stable.
The applicability range is wide enough to implement the 2D fully-developed model over an interesting range of parameters, to retrieve heat transfer correlations.


\begin{figure}[!h]
    \centering
    \includegraphics[width=0.85\textwidth]{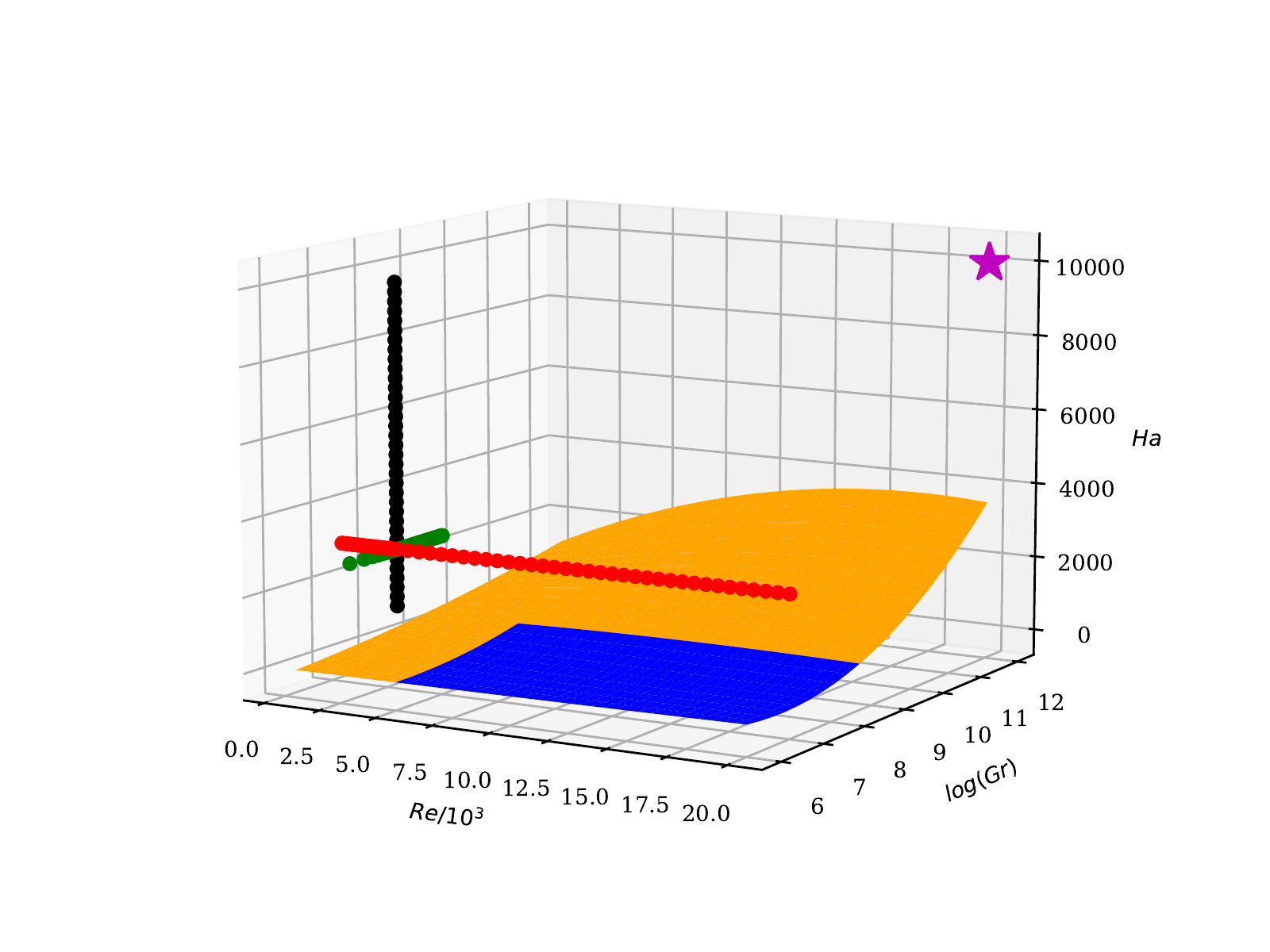}
    \caption{Selected parameters for this study compared to the stability surface provided by Vetcha et al. \cite{Vetcha2013}.}
    \label{fig:rehagrmap}
\end{figure}

 The best fit, as explained in the original reference, corresponds to:

 \begin{equation}\label{eq:ha_cr}
     H\!a_{cr} = P_1\cdot(log\ Gr)^2 + P_2\cdot(log\ Gr) + P_3
 \end{equation}
 with,
 \begin{equation}\label{eq:P1}
     P_1 = -5.98\cdot 10^{-8}\ Re^2+2.284\cdot10^{-3}\ Re + 2.308
 \end{equation}
 \begin{equation}\label{eq:P2}
     P_2 = 1.8277\cdot 10^{-6}\ Re^2-7.3037\cdot10^{-2}\ Re - 22.787
 \end{equation}
 \begin{equation}\label{eq:P3}
     P_3 = -1.37\cdot 10^{-5}\ Re^2+0.57516\ Re -95.8
 \end{equation}

The selected parameters ($Re$, $H\!a$ and $Gr$) for our study are shown together with the stability condition set by Vetcha et al. in Figure \ref{fig:rehagrmap}. 
One can see in blue the stability limit surface by Vetcha et al., in orange the extrapolated surface off range.
The simulated values of $Re$, $Gr$ and $H\!a$ are shown in red, green and black, respectively.
The magenta star represents the DCLL design conditions at DEMO.
This 3D plot shows that the fully-developed model has been applied to the stable region.

The maximum $Ly=H\!a^2/Gr^{0.5}$ (considered to assess the balance of electromagnetic forces to buoyant forces \cite{Lykoudis1990}) for this range of cases is 10\textsuperscript{5}, while the minimum is 225, both well above 1, which suggests flow stability.
The maximum $Re^2/Gr$ (criterion to transition from forced to natural convection \cite{Sparrow1959}) is 400 while the minimum is 0.01.
The nominal conditions for the DCLL are $Ly$~=~10\textsuperscript{2} and $Re^2/Gr$~=~10\textsuperscript{-4}. 
Although the minimum $Re^2/Gr$ in our range of cases is significantly higher than the operating conditions and is inside the stable region provided by \cite{Vetcha2013}, it is still small enough that their results must be used with caution.

\section{Results}

\begin{figure}[!h]
    \centering
    \includegraphics[width=0.75\textwidth]{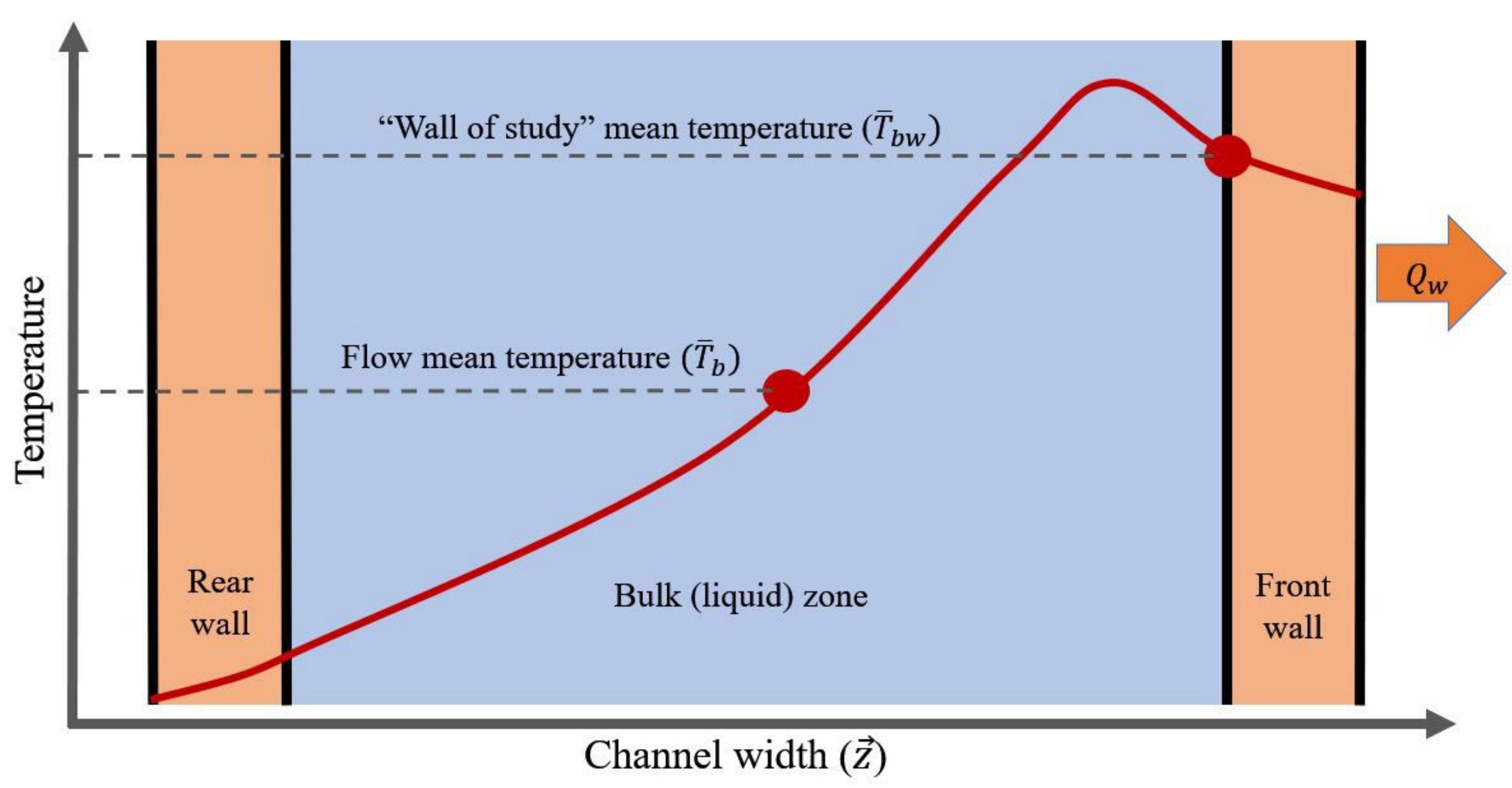}
    \caption{Sketch of a channel center line temperature distribution that causes negative $N\!u$ numbers.}
    \label{fig:NegativeNusselt}
\end{figure}

Due to the exponential heat deposition profile and the heat extraction at the hottest wall, the temperature distribution in the centreline of the channel shows a profile similar to the shown in Figure \ref{fig:NegativeNusselt}. 
Reader should note that even though the heat flux goes from the liquid region to the front wall, the mean temperature of such wall is higher than the mean temperature of the liquid region. 
As indicated at the end of Section \ref{section:Methodology}, the definition of the $N\!u$ number used in this work has been the common definition for heat transferred in pipes and channels. 
Usually in those situations where the heat is extracted from the liquid region through walls, and with no heat generation, the $N\!u$ number is positive, yet the $\Delta T_{bw} = \overline{T}_b-\overline{T}_{bw}>0$. In the case modelled here and shown in Figure \ref{fig:NegativeNusselt}, the heat generation provokes such temperature profile that $\Delta T_{bw} = \overline{T}_b-\overline{T}_{bw}<0$, providing negative $N\!u$ numbers in all results.

\subsection{Nusselt vs. aspect ratio}
\begin{figure}[!h]
    \centering
    \includegraphics[width=0.55\textwidth]{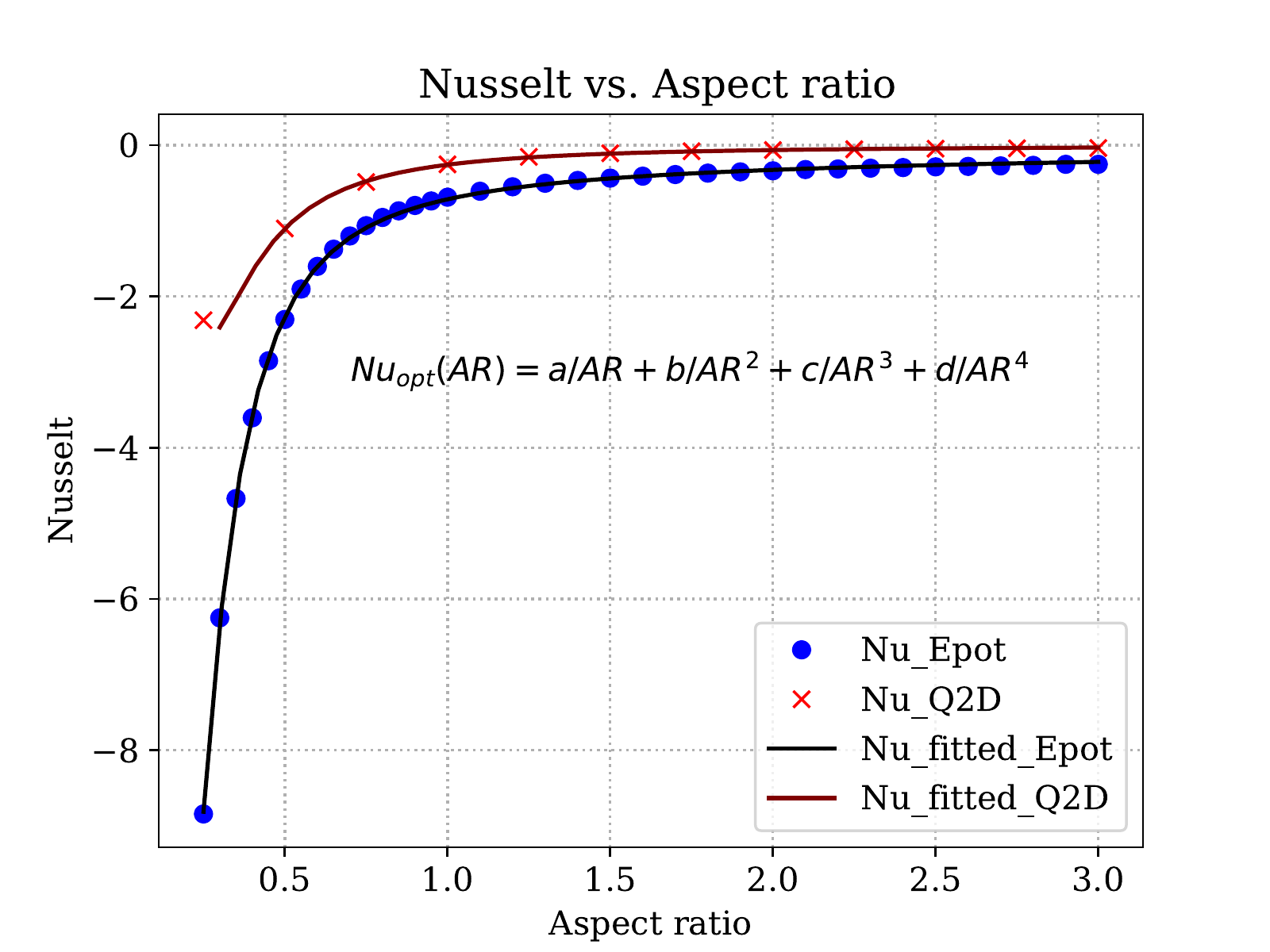}
    \caption{Nusselt number as a function of the aspect ratio}
    \label{fig:Nu_AR}
\end{figure}

The relationship between the Nusselt number and the aspect ratio can be observed in Figure \ref{fig:Nu_AR}. The best fit for the obtained data is:
 \begin{equation}\label{eq:Nu_AR}
     N\!u_{opt}(AR) = a/AR + b/AR^2 + c/AR^3 + d/AR^4
 \end{equation}

The coefficients to fit the function are shown in Table \ref{tab:Nu_AR}.

\begin{table}[!h]
    \centering
    \caption{Fitting parameters for Nusselt number as a function of AR}
    \begin{tabular}{c c c c c}
    \toprule
       Code & a  &  b  &  c  &  d   \\
    \midrule
       Epot & -0.7366  &  0.3295  & -0.3530 & 0.0447    \\
       Q2D  & -0.0477  &  -0.1025 & -0.1422 & 0.0337    \\       
    \bottomrule
    \end{tabular}

    \label{tab:Nu_AR}
\end{table}

\subsection{Nusselt vs. wall conductance ratio}
\begin{figure}[!h]
    \centering
    \includegraphics[width=0.55\textwidth]{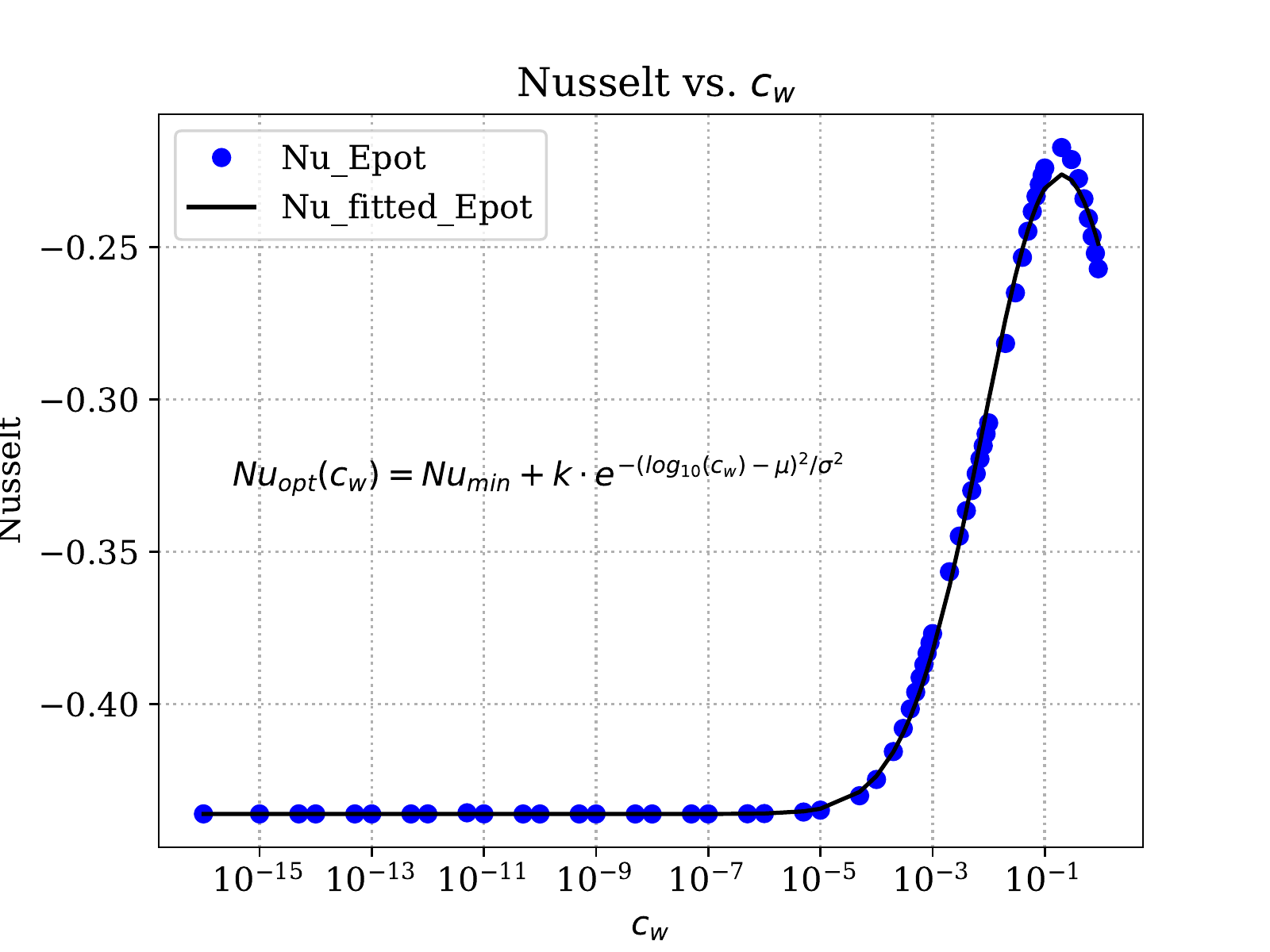}
    \caption{Nusselt number as a function of the wall conductance ratio ($c_w$)}
    \label{fig:Nu_cw}
\end{figure}

The relationship between the Nusselt number and the wall conductance ratio ($c_w)$ can be observed in Figure \ref{fig:Nu_cw}. The best fit for the obtained data is:
 \begin{equation}\label{eq:Nu_cw}
     N\!u_{opt}(c_w) = N\!u_{min} + k\cdot e^{-(log_{10}(c_w)-\mu)^2/\sigma^2}
 \end{equation}

The coefficients to fit the function are shown in Table \ref{tab:Nu_cw}. Note that the Q2D code is not able to simulate different electric conductivities of the wall, since the model considers them electrically insulated.

\begin{table}[!h]
    \centering
    \caption{Fitting parameters for Nusselt number as a function of the wall conductance ratio ($c_w$)}
    \begin{tabular}{c c c c c }
    \toprule
       Code & $N\!u_{min}$ & k  &  $\mu$  &  $\sigma$   \\
    \midrule
       Epot  & -0.4361  &  0.2100 & -0.7089 & 1.9576  \\
    \bottomrule
    \end{tabular}

    \label{tab:Nu_cw}
\end{table}

\subsection{Nusselt vs. Grashof ratio}
\begin{figure}[!h]
    \centering
    \includegraphics[width=0.55\textwidth]{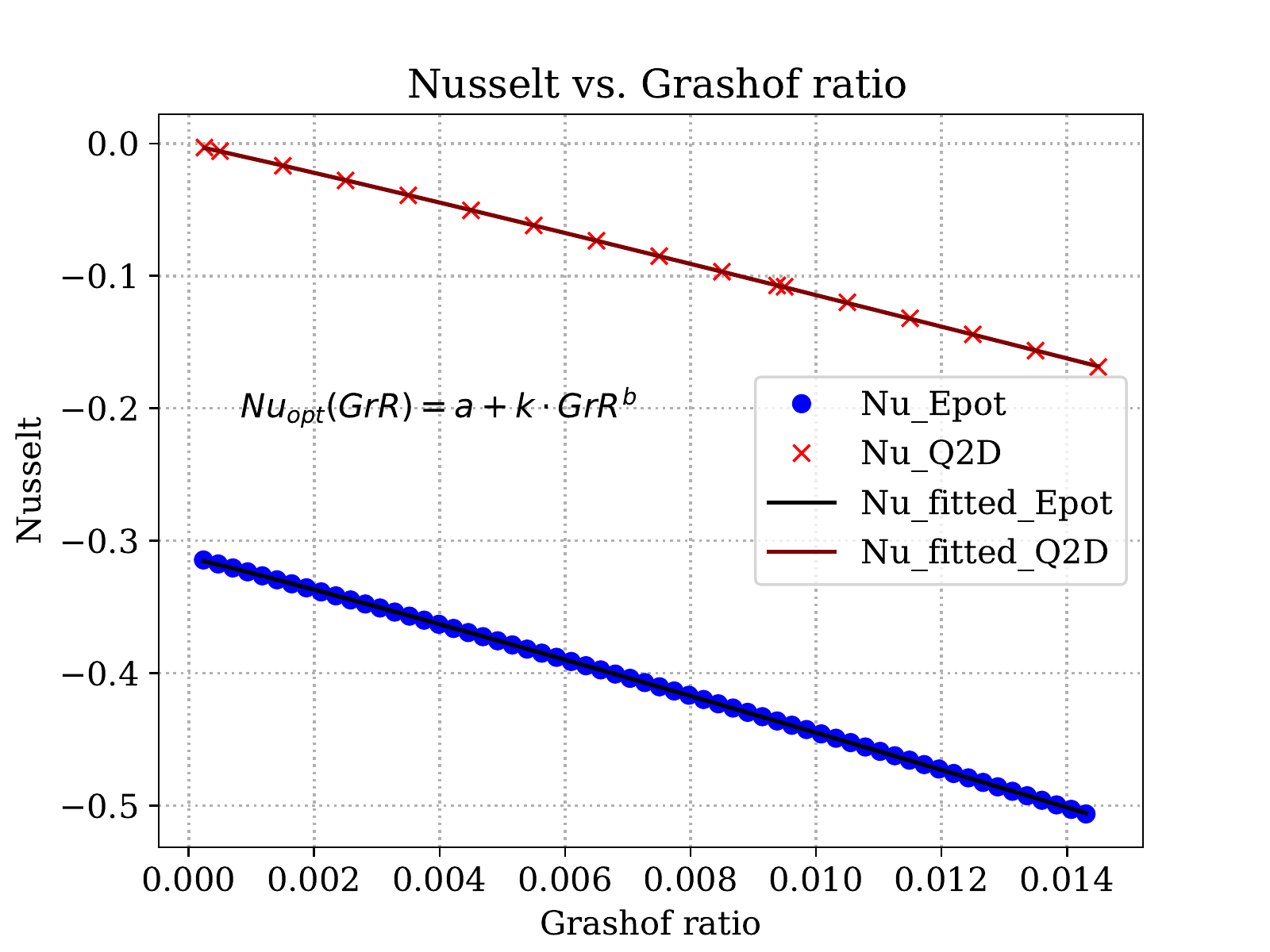}
    \caption{Nusselt number as a function of the Grashof ratio}
    \label{fig:Nu_grr}
\end{figure}

The relationship between the Nusselt number and the Grashof ratio can be observed in Figure \ref{fig:Nu_grr}. The best fit for the obtained data is:
 \begin{equation}\label{eq:Nu_grr}
     N\!u_{opt}(GrR) = a + k\cdot GrR^b 
 \end{equation}

The coefficients to fit the function are shown in Table \ref{tab:Nu_grr}.
Note that the definition given for \textit{GrR} implies that when it is zero, the Nusselt number, which is directly proportional to $Q_W$, must also be zero.
This fact is clearly seen in the results provided by the Q2D code.
For this reason, the value of $a$ has been set to zero in Table~\ref{tab:Nu_grr}.

\begin{table}[!h]
    \centering
    \caption{Fitting parameters for Nusselt number as a function of the Grashof ratio}
    \begin{tabular}{c c c c }
    \toprule
       Code &   k        &   a       &   b    \\
    \midrule
       Epot &   -17.1196  &  -0.3128  &   1.0561\\
    Q2D &  -13.1842 & 0    &  1.0304\\  
    \bottomrule
    \end{tabular}

    \label{tab:Nu_grr}
\end{table}

\subsection{Nusselt vs. exponential coefficient (m)}
\begin{figure}[!h]
    \centering
    \includegraphics[width=0.55\textwidth]{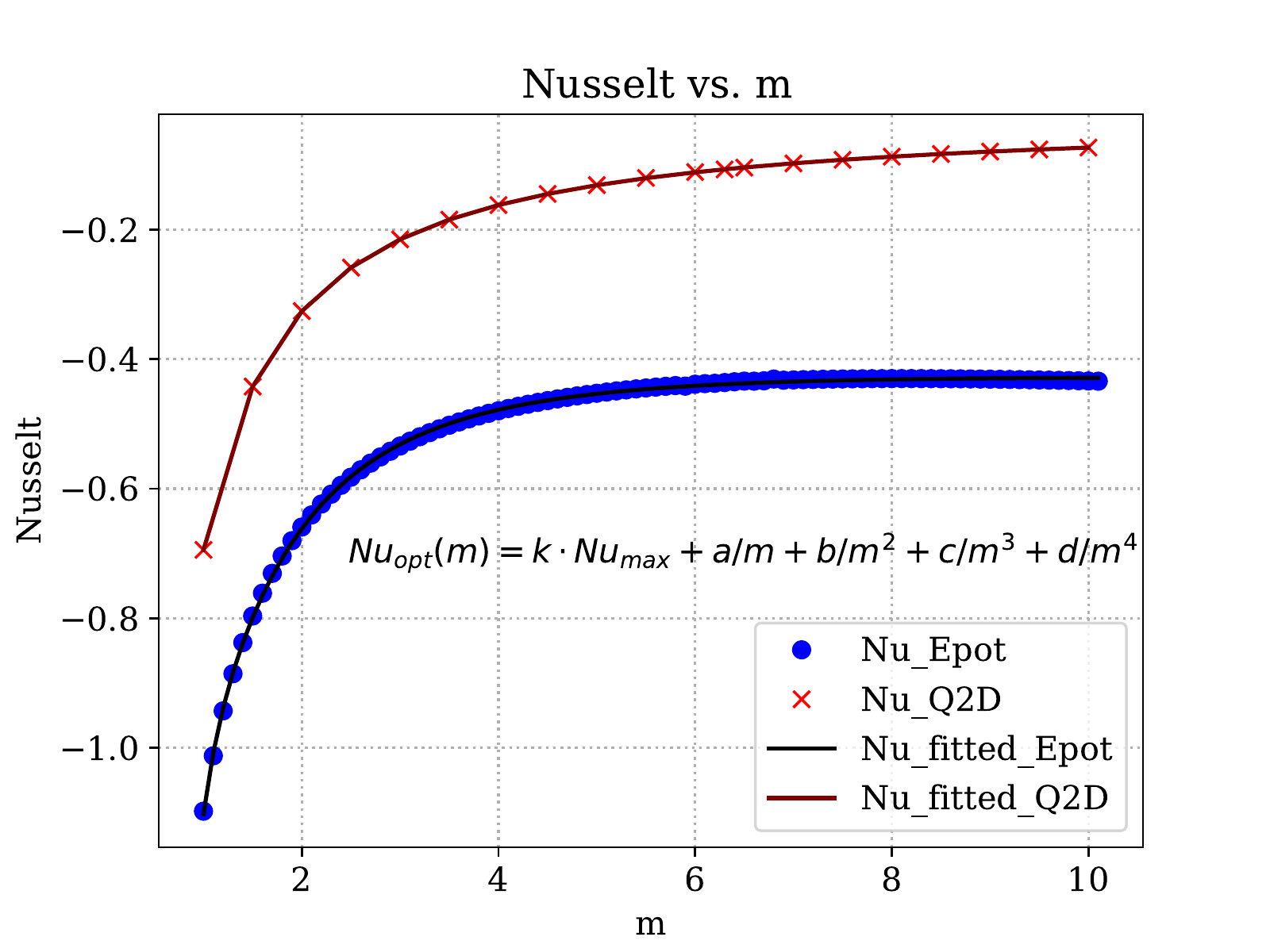}
    \caption{Nusselt number as a function of the exponential coefficient (m)}
    \label{fig:Nu_m}
\end{figure}

The relationship between the Nusselt number and the volumetric heat generation exponential shape coefficient ($m$) can be observed in Figure \ref{fig:Nu_m}. The best fit for the obtained data is:
 \begin{equation}\label{eq:Nu_m}
     N\!u_{opt}(m) = k \cdot N\!u_{max} + a/m + b/m^2 + c/m^3 + d/m^4
 \end{equation}

The coefficients to fit the function are shown in Table \ref{tab:Nu_m}.

\begin{table}[!h]
    \centering
    \caption{Fitting parameters for Nusselt number as a function of $m$}
    \begin{tabular}{c c c c c c c}
    \toprule
       Code &   k       &   $N\!u_{max}$   &    a      &   b      &       c   &  d   \\
    \midrule
       Epot &   1.0499  &    -0.4301     &   0.5269  &  -3.3590 &   3.6777  & -1.4963    \\
       Q2D  &   0.2816  &    -0.0734     &  -0.4954  &  -0.3506 &   0.3084  & -0.1362    \\       
    \bottomrule
    \end{tabular}

    \label{tab:Nu_m}
\end{table}

\subsection{Nusselt vs. Grashof number}
\begin{figure}[!h]
    \centering
    \includegraphics[width=0.55\textwidth]{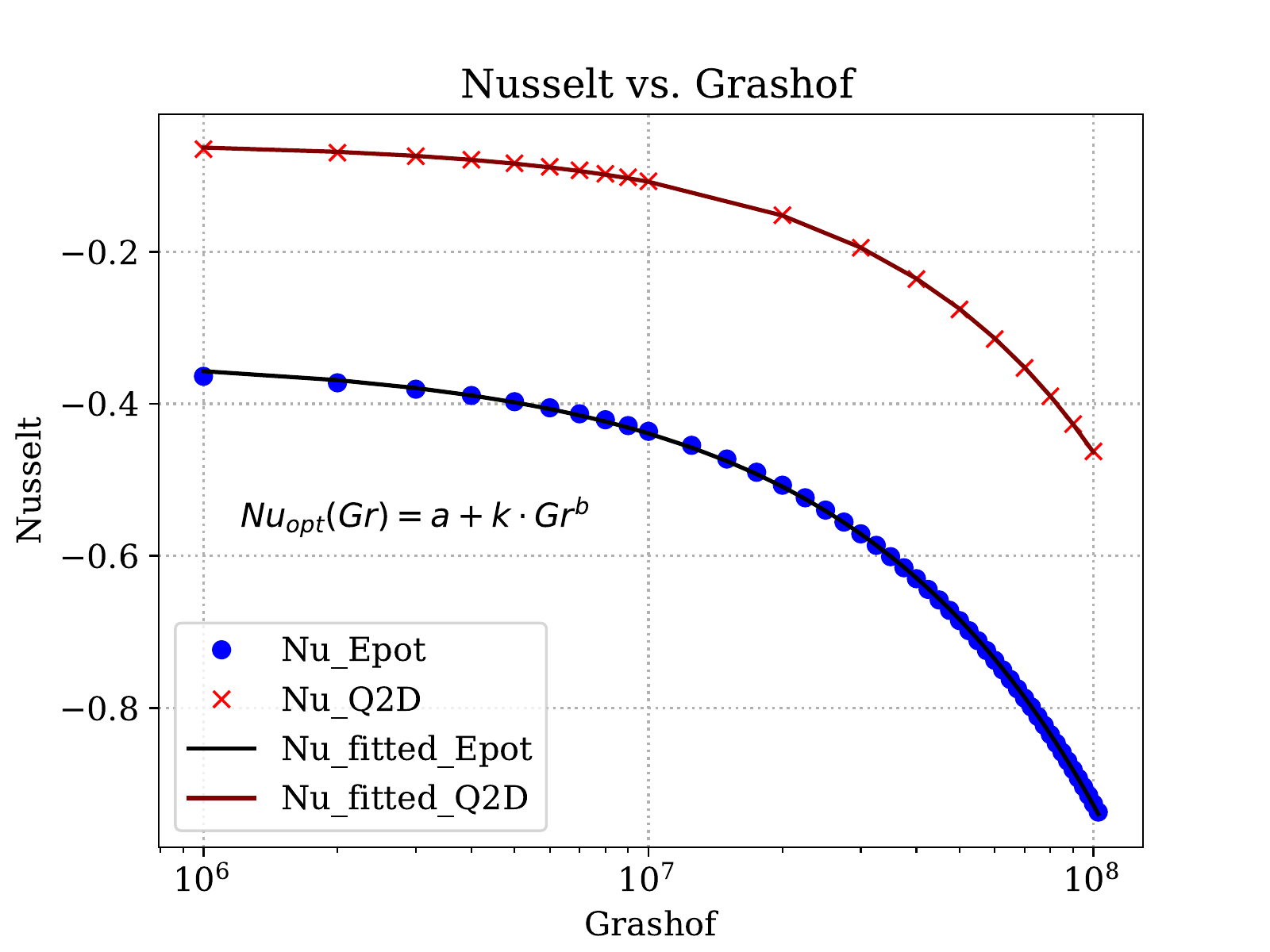}
    \caption{Nusselt number as a function of the Grashof number}
    \label{fig:Nu_gr}
\end{figure}

The relationship between the Nusselt number and the Grashof number can be observed in Figure \ref{fig:Nu_gr}. The best fit for the obtained data is:
 \begin{equation}\label{eq:Nu_gr}
     N\!u_{opt}(Gr) = a + k \cdot Gr^b
 \end{equation}

The coefficients to fit the function are shown in Table \ref{tab:Nu_gr}.

\begin{table}[!h]
    \centering
    \caption{Fitting parameters for Nusselt number as a function of $Gr$}
    \begin{tabular}{c c c c }
    \toprule
       Code &   k        &   a       &   b    \\
    \midrule
       Epot &   -3.5771  &  -0.3407  &  0.7769 \\
       Q2D  &   -2.6466  &  -0.0564  &  0.8984 \\    
    \bottomrule
    \end{tabular}

    \label{tab:Nu_gr}
\end{table}

\subsection{Nusselt vs. Reynolds number}
\begin{figure}[!ht]
    \centering
    \includegraphics[width=0.55\textwidth]{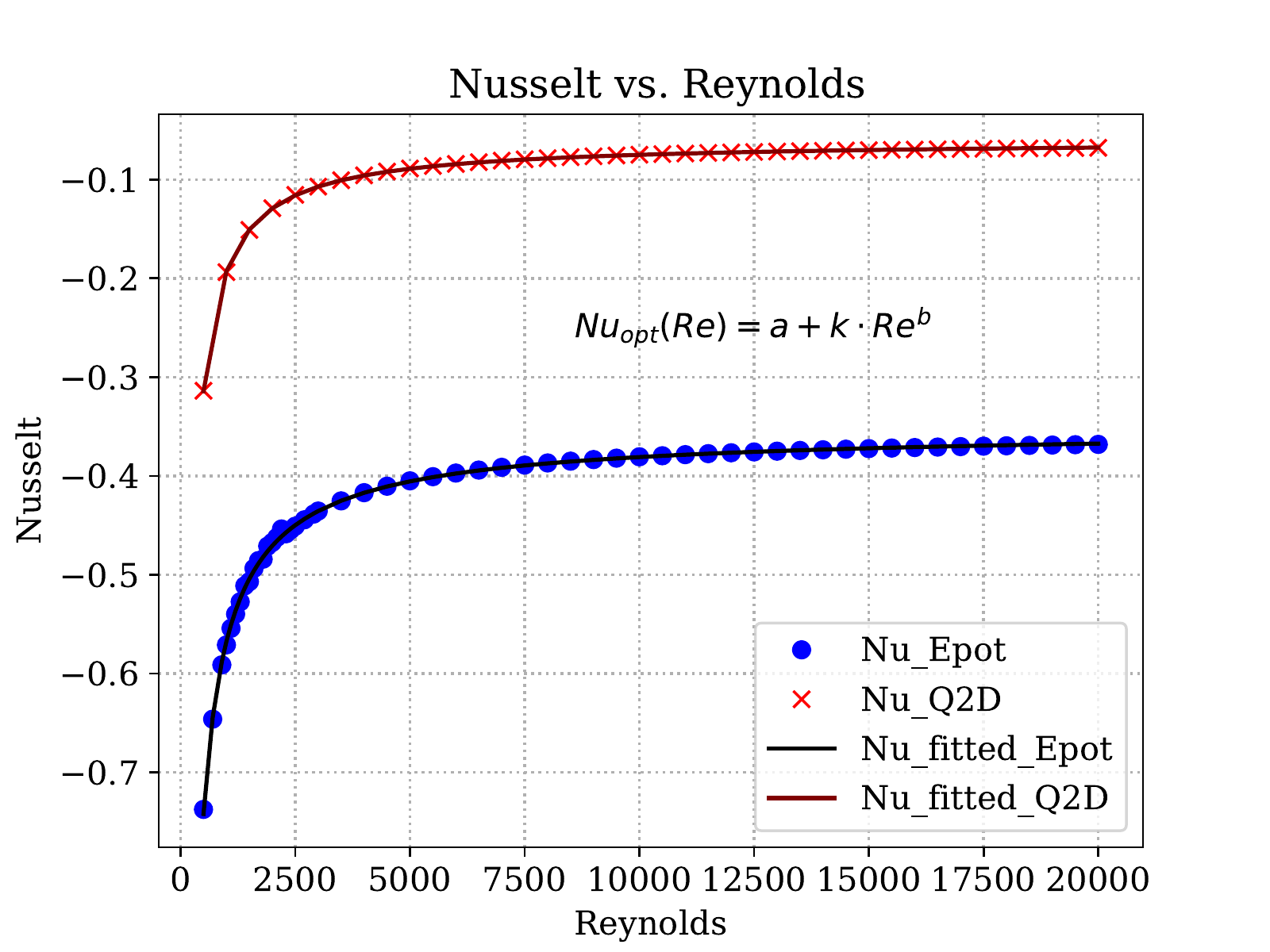}
    \caption{Nusselt number as a function of the Reynolds number}
    \label{fig:Nu_re}
\end{figure}

The relationship between the Nusselt number and the Reynolds number can be observed in Figure \ref{fig:Nu_re}. The best fit for the obtained data is:
 \begin{equation}\label{eq:Nu_re}
     N\!u_{opt}(Re) = a + k \cdot Re^b
 \end{equation}

The coefficients to fit the function are shown in Table \ref{tab:Nu_re}.

\begin{table}[h]
    \centering
    \caption{Fitting parameters for Nusselt number as a function of $Re$}
    \begin{tabular}{c c c c }
    \toprule
       Code &   k        &   a       &   b    \\
    \midrule
       Epot &   -77.7624  &  -0.3503  &  -0.8513 \\
       Q2D  &   -85.7331  &  -0.0596  &  -0.9364 \\
    \bottomrule
    \end{tabular}

    \label{tab:Nu_re}
\end{table}

\subsection{Nusselt vs. Hartmann number}
\begin{figure}[!ht]
    \centering
    \includegraphics[width=0.55\textwidth]{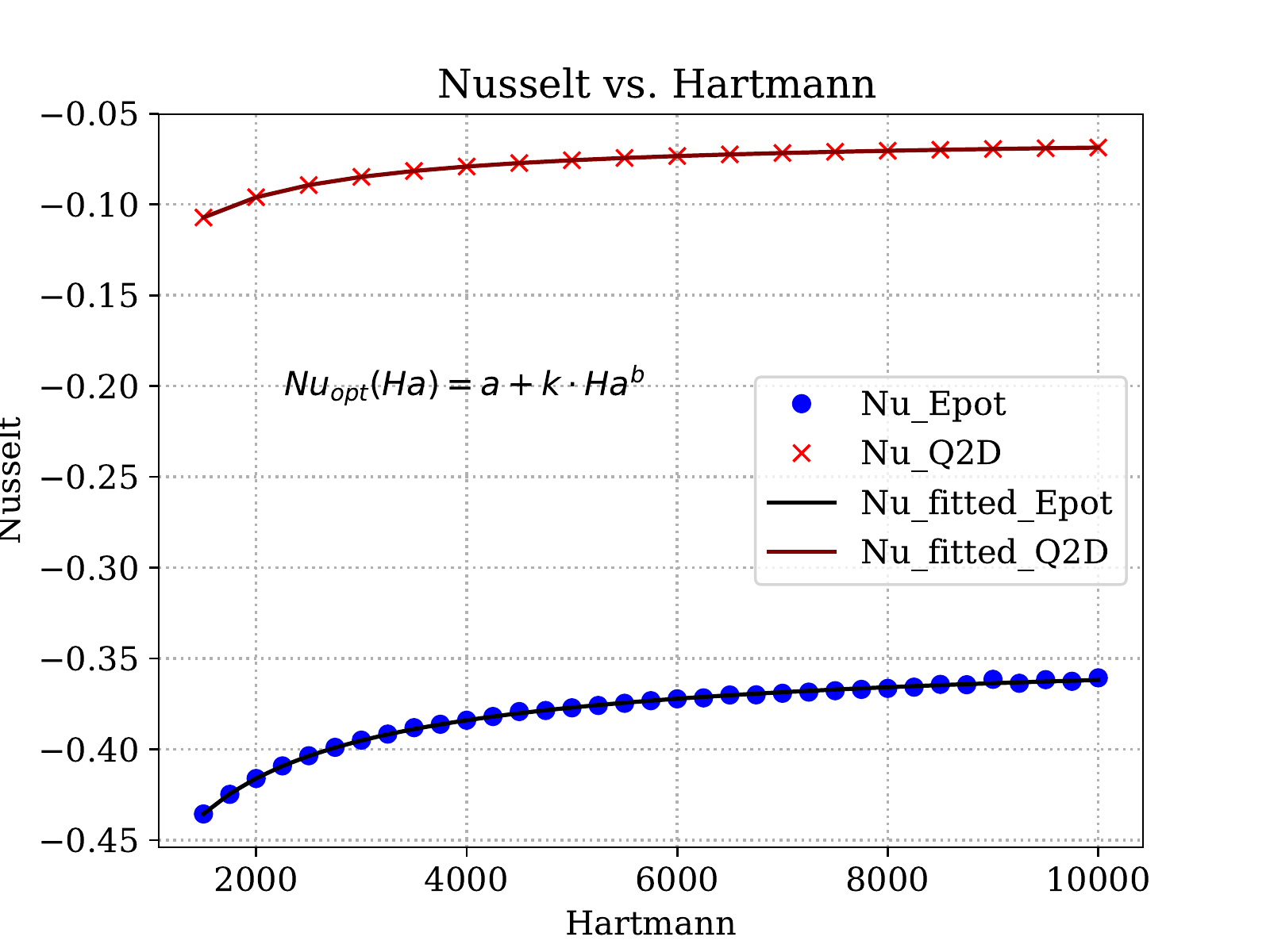}
    \caption{Nusselt number as a function of the Hartmann number}
    \label{fig:Nu_ha}
\end{figure}

The relationship between the Nusselt number and the Hartmann number can be observed in Figure \ref{fig:Nu_ha}. The best fit for the obtained data is:
 \begin{equation}\label{eq:Nu_ha}
     N\!u_{opt}(H\!a) = a + k \cdot H\!a^b
 \end{equation}

The coefficients to fit the function are shown in Table \ref{tab:Nu_ha}.

\begin{table}[h]
    \centering
    \caption{Fitting parameters for Nusselt number as a function of $H\!a$}    
    \begin{tabular}{c c c c}
    \toprule
       Code &   k        &   a       &   b    \\
    \midrule
       Epot &   -36.0443 &  -0.3418  &  -0.8139 \\
       Q2D  &   -53.6212 &  -0.06138  & -0.9660 \\    
          
    \bottomrule
    \end{tabular}

    \label{tab:Nu_ha}
\end{table}

\section{Discussion}
The dependence of the heat transfer coefficient (through the $N\!u$ number) on the selected seven dimensionless parameters has been correlated. Almost 140 simulations of the 2D fully-developed model with the electric potential code (\textit{Epot}) have been run in parallel in the HPC Marconi. 
All simulations run with the Q2D model could be solved in a personal computer in series.

The selected correlation models fit well the obtained data. 
The parametric study has been performed varying one variable from the base case each time.

The comparison between codes yields a very interesting conclusion. 
The influence of each variable to the heat transfer is well captured by both codes. 
The difference in all cases but $AR$ is an absolute value associated with the fact that the mean temperature in the wall is different in each code. 
Therefore, the correlated data using one code is approximately parallel to the correlated data with the other code.

Regarding the $AR$, the influence of the heat transferred in the Hartmann walls and the channel corners is modified with the $AR$, what prevents the correlated data to be parallel.

An interesting Nusselt peak has been identified using the electric potential code around $c_w \sim 0.2$.
It corresponds to the maximum velocity of the jets formed in the side boundary layer. 
The velocity peak strongly reduces the $\nabla \theta_f$ but also a bit the $\Delta T_{bw}$. 
The combination of effects results in a higher $N\!u$ number.

The most surprising result of this work is that the best fit for the dimensionless variables $H\!a$, $Re$, $Gr$, and $GrR$ is an equation of similar form. 
Rearranging the relationship between them, and keeping aside the influence of $c_w$, $m$, $AR$, the following expressions are proposed for the two models:

\begin{equation}\label{eq:Nu_ha_re_gr_grr_Epot}
   N\!u_{opt\ Epot}(H\!a,Re,Gr,GrR) = r + \left(s + k \cdot \left(\frac{Gr}{Re \cdot H\!a}\right)^a\right)\cdot GrR^d
\end{equation}

\begin{equation}\label{eq:Nu_ha_re_gr_grr_Q2D}
   N\!u_{opt\ Q2D}(H\!a,Re,Gr,GrR) = \left(s + k \cdot \left(\frac{Gr}{Re \cdot H\!a}\right)^a\right)\cdot GrR^d
\end{equation}

The fitting coefficients are shown in Table \ref{tab:Nu_ha_re_gr_grr}.
The coefficients provide the
best fit using the results obtained from the parametric analysis in which only one
parameter varies.
They have been slightly rounded to couple the ratios between $H\!a$, $Re$, $Gr$ into one term.

\begin{table}[!h]
  \centering
  \caption{Fitting parameters for Nusselt number as a function of $H\!a$, $Re$, $Gr$ and $GrR$ in the single parameter variation study}
  \begin{tabular}{c c c c c c}
  \toprule
    Code  &  r		&	s    &  k    &    a   &   d\\
  \midrule
    Epot  & -0.3125 &	-1.89 &  -3  &    0.8 &  1.05 \\
    Q2D   & 	-	&	-7.71 & -3.34&    0.9 &  1.05 \\    
  \bottomrule
  \end{tabular}
  \label{tab:Nu_ha_re_gr_grr}
\end{table}

The resulting RMSE using the expression and coefficients proposed for the \textit{Epot} results is 0.00233.
The RMSE for the proposed Q2D correlation and coefficients is 0.0012.

\begin{figure}[!ht]
  \centering
  \includegraphics[width=0.85\textwidth]{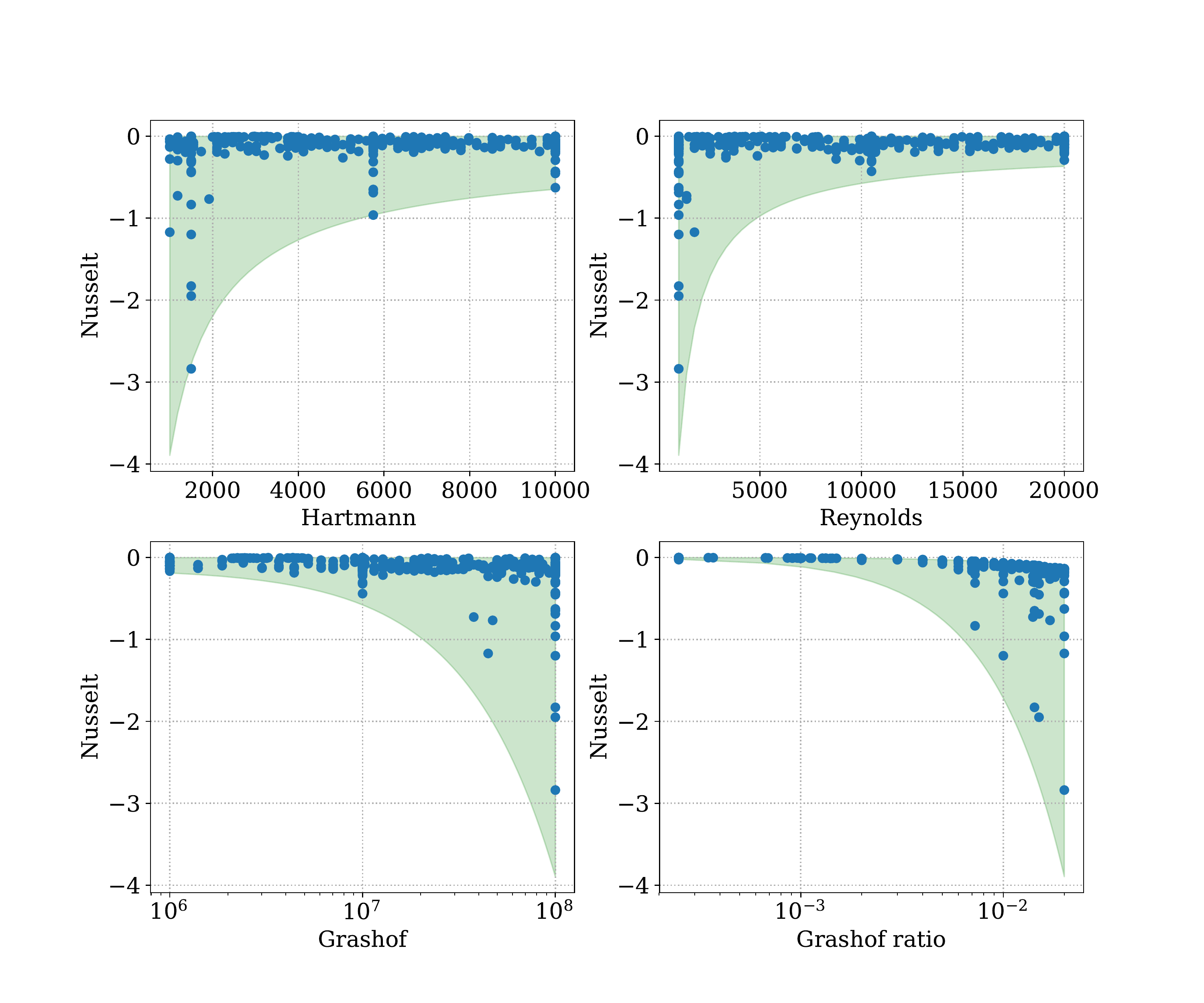}
  \caption{Nusselt number as a function of $H\!a$, $Re$, $Gr$ and $GrR$ between the maximum and minimum of the fitting curve (\ref{eq:Nu_ha_re_gr_grr_Q2D_separats})}
  \label{fig:Nucorr}
\end{figure}

Running the Q2D code for a fully-developed case confirmed that it effectively predicts the influence of the dimensionless parameters on the flow profile.
Following these encouraging results and given the fast and lightweight calculation properties of this numerical approach, a final set of cases was run to confirm the accuracy of the proposed correlation and coefficients.
In this final set of cases, the dimensionless parameters could be any value between the minimum and maximum of the studied range.
Consequently, this approach varies several parameters from one simulation to another.

\begin{figure}[!ht]
  \centering
  \includegraphics[width=0.85\textwidth]{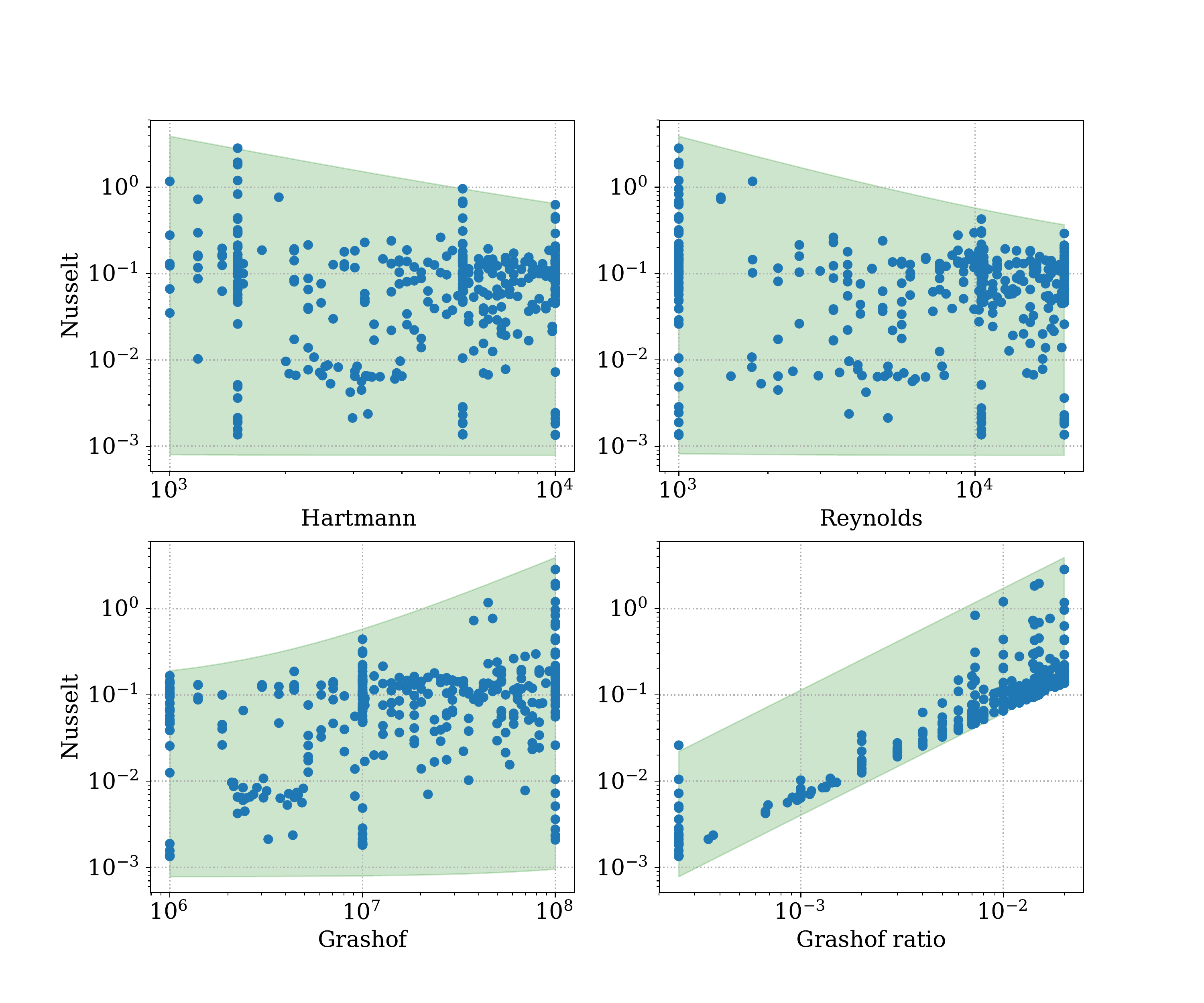}
  \caption{Absolute Nusselt number as a function of $H\!a$, $Re$, $Gr$ and $GrR$ in logarithmic scale between the maximum and minimum of the fitting curve (\ref{eq:Nu_ha_re_gr_grr_Q2D_separats})}
  \label{fig:Nucorr_log}
\end{figure}

The dependence of the Nusselt number on the four dimensionless parameters is shown in Figures~\ref{fig:Nucorr} and~\ref{fig:Nucorr_log}.
Figure~\ref{fig:Nucorr_log} shows the logarithm of the absolute value of the Nusselt number to provide more detail on the obtained $N\!u$ when $GrR$ tends to zero.
The obtained results are represented with blue dots, while the green bands represent the range between the maximum and minimum possible $N\!u$ obtained with the proposed correlation (\ref{eq:Nu_ha_re_gr_grr_Q2D_separats}) using all possible combinations of dimensionless variables in the studied range.
The fitting coefficients for this function have been updated and are shown in Table~\ref{tab:Corr_comparison}.

\begin{equation}\label{eq:Nu_ha_re_gr_grr_Q2D_separats}
   N\!u_{opt\ Q2D}(H\!a,Re,Gr,GrR) = \left(s + k \cdot Gr^a \cdot Re^b \cdot H\!a^c \right)\cdot GrR^d
\end{equation}

The effect of assigning an independently variable exponent to each parameter $Re$, $H\!a$ and $Gr$ as in (\ref{eq:Nu_ha_re_gr_grr_Q2D_separats}), rather than assigning the same exponent $a$ to all three (\ref{eq:Nu_ha_re_gr_grr_Q2D}), was assessed using the results of this study.
The data obtained allowed different fitting functions to be compared.
The selected fitting functions for comparison were: 
\begin{itemize}
	\item The one based on a single-variable parametric analysis (\ref{eq:Nu_ha_re_gr_grr_Q2D}).
	\item The same function (\ref{eq:Nu_ha_re_gr_grr_Q2D}) using all the data from this last multivariable set of cases to adjust the fitting coefficients.
	\item A new function using all of the data from this last multi-variable set of cases, separating the exponents of $Re$, $H\!a$ and $Gr$ (\ref{eq:Nu_ha_re_gr_grr_Q2D_separats}).
\end{itemize}

A comparison of the RMSE of each of the fitting functions for this set of cases is shown in Table~\ref{tab:Corr_comparison}.

\begin{table}[!ht]
  \centering
  \caption{Fitting parameters for Nusselt number as a function of $H\!a$, $Re$, $Gr$ and $GrR$ using a Q2D fully-developed model for different set of cases and functions}
  \begin{tabular}{c c c c c c c c c}
  \toprule
    Function &	Set of cases &	RMSE	& s   &	k  & a  &  b  &  c 	 & 	d	\\
  \midrule
    (\ref{eq:Nu_ha_re_gr_grr_Q2D}) &	Single var.	 		 & 0.0194	& -7.71 & -3.34 & 0.9 & -  &  -  & 1.05	\\
	  (\ref{eq:Nu_ha_re_gr_grr_Q2D}) &	Multiple var. 		 & 	0.0087	& -13.50 & -5.82 & 0.91 & -  &  -  & 1.18 \\
    (\ref{eq:Nu_ha_re_gr_grr_Q2D_separats})&Multiple var. & 0.0063	& -13.95 & -3.46 & 0.93 & -0.93 & -0.87 & 1.18 \\    
  \bottomrule
  \end{tabular}
  \label{tab:Corr_comparison}
\end{table}

The first two rows of the table show that when the number of sample cases is increased, the resulting coefficients are more accurate, providing a lower RMSE.
As expected, the function that treats the exponents independently (\ref{eq:Nu_ha_re_gr_grr_Q2D_separats}) shows a lower RMSE than the functions that gather them into one term (\ref{eq:Nu_ha_re_gr_grr_Q2D}), but the difference is small.
The exponent a = 0.91 and the ratio $Gr/(H\!a \cdot Re)$ are the most important findings of this study.

\section*{Acknowledgments}
This work has been carried out within the framework of the EUROfusion Consortium, funded by the European Union via the Euratom Research and Training Programme (Grant Agreement No 101052200 — EUROfusion). Views and opinions expressed are however those of the author(s) only and do not necessarily reflect those of the European Union or the European Commission. Neither the European Union nor the European Commission can be held responsible for them.

The authors thank as well the contribution of Asso\-cia\-ció/Col·le\-gi d'En\-gi\-nyers In\-dus\-trials de Ca\-ta\-lu\-nya with Fun\-da\-ció Cai\-xa d'En\-gi\-nyers' financial support.

The authors would like to acknowledge either the work done by Eduardo Iraola and Cristina Lampón in the previous work \cite{Suarez2021} as well as the contribution of Badreddine Satouri in the initial phase of data analysis, during his internship.

\printbibliography

\end{document}